\documentclass[
amsmath, amssymb, 
aps, pra, reprint,
twocolumn,
superscriptaddress,
floatfix,
showkeys,
nofootinbib
]{revtex4-1}

\usepackage[unicode]{hyperref}

\usepackage{graphicx} 
\usepackage{float}
\usepackage{wrapfig}
\usepackage{array}
\usepackage{diagbox}
\usepackage{chngpage}
\usepackage{subfigure}
\usepackage{multirow}
\usepackage{tabularx}
\usepackage{hhline}
\usepackage[table]{xcolor}
\usepackage{slashbox}

\usepackage{braket}
\usepackage{accents}
\usepackage{upgreek}
\usepackage{color}

\usepackage{csquotes}
\usepackage{datetime}
\usepackage{units}
\usepackage{siunitx}

\usepackage{appendix}
\usepackage{mathtools}

\usepackage{textcomp}

\newcommand{\Yb}{{}^{171}\textrm{Yb}^{+}}
\newcommand{\ion}[2]{{}^{#2}\textrm{#1}^{+}}
\newcommand{\LS}[2]{{}^2{\textrm{#1}}_{#2}}
\newcommand{\LSB}[4]{{}^{#1}{\textrm{#2}}[#3]_{#4}}

\begin{document}

\title{Scalable hyperfine qubit state detection via electron shelving in the \texorpdfstring{${}^2$D$_{5/2}$}{2D5/2} and \texorpdfstring{${}^2$F$_{7/2}$}{2F7/2} manifolds in \texorpdfstring{$\Yb$}{171Yb+}}

	\affiliation{
	ARC Centre of Excellence for Engineered Quantum Systems, The University of Sydney, School of Physics, NSW, 2006, Australia }
	\affiliation{
    The University of Sydney Nano Institute (Sydney Nano), The University of Sydney, NSW 2006, Australia	}

\author{C. L. Edmunds}
\thanks{edmunds.claire@gmail.com}
    \affiliation{
	ARC Centre of Excellence for Engineered Quantum Systems, The University of Sydney, School of Physics, NSW, 2006, Australia }
\author{T. R. Tan}
	\affiliation{
	ARC Centre of Excellence for Engineered Quantum Systems, The University of Sydney, School of Physics, NSW, 2006, Australia }
\author{A. R. Milne}
	\affiliation{
	ARC Centre of Excellence for Engineered Quantum Systems, The University of Sydney, School of Physics, NSW, 2006, Australia }
\author{A. Singh}
	\thanks{
	current address: Department of Physics, University of California, Berkeley, California 94720, USA}
	\affiliation{
	ARC Centre of Excellence for Engineered Quantum Systems, The University of Sydney, School of Physics, NSW, 2006, Australia }
\author{M. J. Biercuk}
	\affiliation{
	ARC Centre of Excellence for Engineered Quantum Systems, The University of Sydney, School of Physics, NSW, 2006, Australia }
     \affiliation{Q-CTRL Pty Ltd, Sydney, NSW, 2006, Australia}
\author{C. Hempel}
\thanks{cornelius.hempel@gmail.com}
	\affiliation{
	ARC Centre of Excellence for Engineered Quantum Systems, The University of Sydney, School of Physics, NSW, 2006, Australia }
	\affiliation{
    The University of Sydney Nano Institute (Sydney Nano), The University of Sydney, NSW 2006, Australia	}

\date{\today}

\begin{abstract}
    Qubits encoded in hyperfine states of trapped ions are ideal for quantum computation given their long lifetimes and low sensitivity to magnetic fields, yet they suffer from off-resonant scattering during detection often limiting their measurement fidelity. In $\Yb$ this is exacerbated by a low fluorescence yield, which leads to a need for complex and expensive hardware -- a problematic bottleneck especially when scaling up the number of qubits. 
    We demonstrate a detection routine based on electron shelving to address this issue in $\Yb$ and achieve a 5.6$\times$ reduction in single-ion detection error on an avalanche photodiode to \num{1.8(2)e-3} in a 100~$\upmu$s detection period, and a 4.3$\times$ error reduction on an electron multiplying CCD camera, with \num{7.7(2)e-3} error in 400~$\upmu$s. 
    We further improve the characterization of a repump transition at
    760~nm to enable a more rapid reset of the auxiliary $\LS{F}{7/2}$ states populated after shelving. 
    Finally, we examine the detection fidelity limit using the long-lived $\LS{F}{7/2}$ state, achieving a further 300$\times$ and 12$\times$ reduction in error to \num{6(7)e-6} and \num{6.3(3)e-4} in 1~ms on the respective detectors. While shelving-rate limited in our setup, we suggest various techniques to realize this detection method at speeds compatible with quantum information processing, providing a pathway to ultra-high fidelity detection in $\Yb$.  
\end{abstract}

\maketitle

\section{Introduction}

Trapped ions have seen a resurgence as a leading platform for the development of quantum information systems. In recent years, a primary area of research  has been the quality of single- and two-qubit gates, where fidelities of better than  99.99\%~\cite{Harty:2014,Gaebler:2016,Sepiol:2019} and 99.9\%~\cite{Gaebler:2016, Ballance:2016} have been reported, respectively. Enabled by quantum control techniques, such as amplitude, frequency and phase-modulation~\mbox{\cite{Choi:2014, Leung:2018, Milne:2020}}, high-fidelity two-qubit gates are now possible at high speeds~\cite{Schafer:2018, Zhang:2020} and also across larger qubit registers~\cite{Wright:2019, Bentley:2020}. Progress in this domain has allowed for the implementation of longer and more complex quantum circuits~\cite[e.g.][]{Nam.2020, Erhard:2020, Egan.2020}. Yet, as the number of qubits in a joint register -- and thereby the potential size of a correlated state -- grows, an increasingly important area for improvement becomes state-detection fidelity. Detection errors are generally statistically independent and scale at least linearly with the number of qubits.  They therefore quickly become a significant factor limiting the overall performance of a multi-qubit register, e.g. in the context of active quantum error correction conditioned on stabilizer measurements~\cite{Schindler.2011,Nigg.2014,Bermudez.2017,Egan.2020}.

Various qubit encodings are available in trapped ions which bring with them different advantages and drawbacks -- including in the area of demonstrated measurement fidelity. One can either choose two ground states of the fine- or hyperfine structure for the encoding, or split the logical states across a ground- and a metastable state to form an optical qubit~\cite{Bruzewicz:2019}.  Hyperfine qubits such as $\ion{Be}{9}$\cite{Gaebler:2016}, $\ion{Ca}{43}$\cite{Benhelm.2008,Kirchmair:2009} or $\Yb$~\cite{Olmschenk:2007} as considered here, are an attractive choice in that they do not suffer from energy relaxation ($T_1$ decay) like optical qubits, and also offer so-called \enquote{clock states} that are first-order insensitive to perturbations from magnetic fields (providing long $T_2$ coherence). Here, the qubit states are separated by microwave frequencies on the order of several to tens of GHz enabling the use of low-noise microwave sources for high-fidelity qubit control~\cite{Harty:2014,Ball:2016, Edmunds:2020}.  

Measurement on either category of trapped-ion qubit is generally performed using state-dependent laser-induced fluorescence~\cite{Wineland.1995}, whereby one logical state, the \enquote{bright state}, scatters photons and the other does not, hence being referred to as \enquote{dark state}.  Optical qubits enable efficient discrimination between these states and have shown high detection fidelities, leveraging the large energy-level separation of the qubit manifold~\cite{Myerson:2008, Burrell:2010}.  By contrast, when using hyperfine qubits the relatively small energy gap between the qubit states results in unwanted off-resonant scattering during detection. This scattering limits the useful duration of the detection period and thereby the number of photons that can be collected in it, negatively impacting the ability to distinguish qubit states from associated photon-detection-probability distributions.  To overcome this obstacle, one may pursue the use of new complex imaging and detection hardware~\cite{Debnath:2016, Seif.2018, Crain:2019, Todaro:2020, Zhukas:2020} or software-based data processing of time-resolved information~\cite{Hume:2007, Myerson:2008, Hemmerling:2012, Wolk:2015, Seif.2018, Ding.2019}.  As qubit numbers are increased, however, the overhead for detection hardware and software can become limiting, motivating an exploration of complementary ``physics-based'' schemes to improve measurement fidelity in hyperfine qubits.  

In this paper, we borrow a detection technique widely employed in optical qubits to perform electron shelving on a hyperfine $\Yb$ qubit to increase detection fidelity without modification of detection hardware or software.  By shelving the population of one of the qubit states to a metastable state separated by an optical transition, we can detect population remaining in the qubit manifold without being limited by off-resonant scattering and the resulting leakage to the other logical state. We implement this method using a quadrupole transition at 411~nm from the $\LS{S}{1/2}$ qubit manifold to the $\LS{D}{5/2}$ state, and separately to the extremely long-lived metastable $\LS{F}{7/2}$ level (Fig.~\ref{fig:full_energy_levels}).  We also employ a repump laser at 760~nm to efficiently restore all population to the qubit manifold after the detection period via the rapidly decaying $\LSB{1}{D}{3/2}{3/2}$ state. In our work, we further combine the shelving routine with efficient software post-processing techniques using photons collected on an avalanche photodiode (APD) and an electron multiplying charged coupled device (EMCCD) camera, using a time-resolved, non-adaptive maximum likelihood protocol on the APD~\cite{Wolk:2015} and a machine-learning-based image classifier on the EMCCD.  We characterize and compare the various routines, demonstrating measurement-fidelity improvements up to $300\times$ leveraging the $\LS{F}{7/2}$ level, and describe how this may be efficiently integrated into quantum information experiments.

\section{Trapped ion qubit state detection}

Various approaches are being pursued to improve qubit state detection with ions, which can be broadly classed as hardware-, software-, or physics-based. The first category uses specialized detectors, such as superconducting nanowire single photon detectors (SNSPDs) either stand-alone~\cite{Crain:2019} or embedded in a surface-electrode RF trap~\cite{Todaro:2020}, multi-channel photomultiplier tubes (PMTs)~\cite{Debnath:2016,Seif.2018}, or fast intensified cameras~\cite{Zhukas:2020}. Several software-based methods have been demonstrated to improve the final state estimation. Combining a record of the incident timing of photons during detection with prior knowledge such as the expected fluorescence rate and decay times $\tau_B$ and $\tau_D$ from the bright and dark states, one can infer the final state from a maximum likelihood calculation~\cite{Myerson:2008}. Furthermore, if real-time data processing is available, the same detection fidelity can be achieved in shorter detection times using an adaptive version of this technique. W{\"o}lk et al.~\cite{Wolk:2015} analyze the time-resolved detection methods for the case of $\Yb$, which had been experimentally demonstrated for optical qubits by Myerson et al.~\cite{Myerson:2008}, and hyperfine qubits by Hume et al.~\cite{Hume:2007} and Hemmerling et al.~\cite{Hemmerling:2012}. Other software-based approaches include recent work by Ding et al.~\cite{Ding.2019} investigating the use of machine-learning methods for state estimation, implemented in hardware on an FPGA with a single $\Yb$ qubit; they achieve similar results to Seif et al.~\cite{Seif.2018}, who apply machine learning methods to the time-resolved readout from a PMT array in post-processing.

\begin{figure}[t!]
    \centering
    \includegraphics[scale=1]{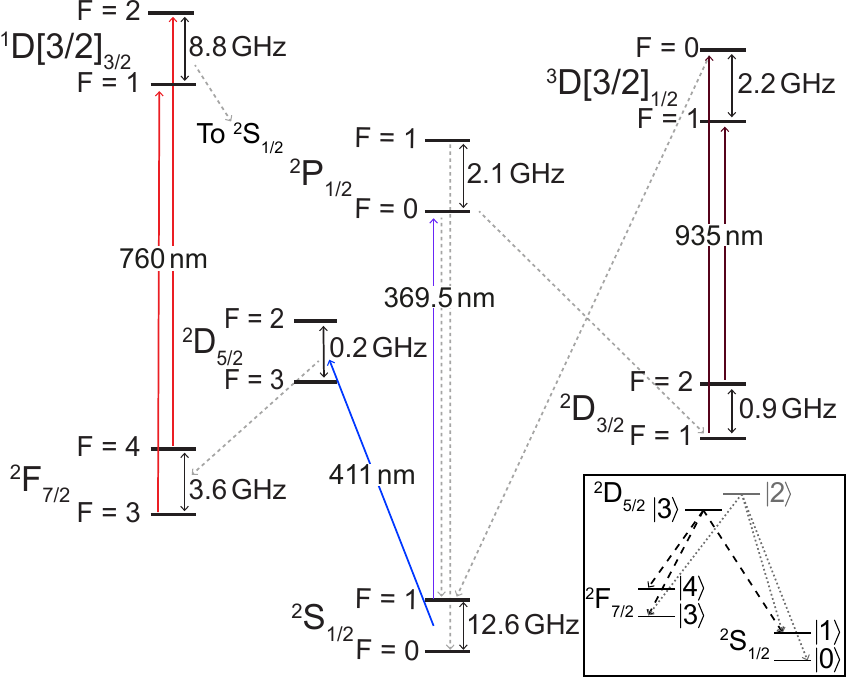}
    \caption{Selected energy levels and laser frequencies for $\Yb$. The qubit levels are encoded in the $\LS{S}{1/2}$ hyperfine manifold, shown at the bottom center. Zeeman levels have been omitted for visual clarity and the dashed gray lines show relevant spontaneous decays. (Inset) Decay channels from $\LS{D}{5/2}$, with a simplified hyperfine notation.}
    \label{fig:full_energy_levels}
\end{figure}

Detection of optical qubits falls under the physics-based approaches, achieving a very high signal-to-noise ratio through what is generally referred to as \enquote{electron shelving} after Hans Dehmelt~\cite{Dehmelt:1975, Nagourney:1986}. In such settings, measurement fidelities of $\geq 99.99\%$ have been reported for an optical qubit encoded in $^{40}$Ca$^+$ using time-resolved measurements of fluorescence~\cite{Myerson:2008}, and separately, without time resolution, on an EMCCD camera~\cite{Burrell:2010}. The detection fidelity of an optical qubit is fundamentally limited by the lifetime of the metastable state (a $T_1$ decay process). Although the associated decay rate during detection is often low under typical detection times, the decay probabilities are independent for each ion and can thereby quickly become the limiting factor to the overall \emph{state} detection fidelity in larger qubit registers. To address this problem, a newer generation of fast cameras is being developed, which allow time-resolved measurements to be carried out while also providing spatial resolution for identification of the unique quantum state in a multi-qubit register. First demonstrations have recently been reported for an electron-shelved read-out in $^{138}$Ba$^+$~\cite{Zhukas:2020}, reaching $\sim99.99\%$ fidelity for a single qubit and at least 99.7\% in a four-qubit register.

\begin{figure}[t]
    \centering
    \includegraphics[scale=1]{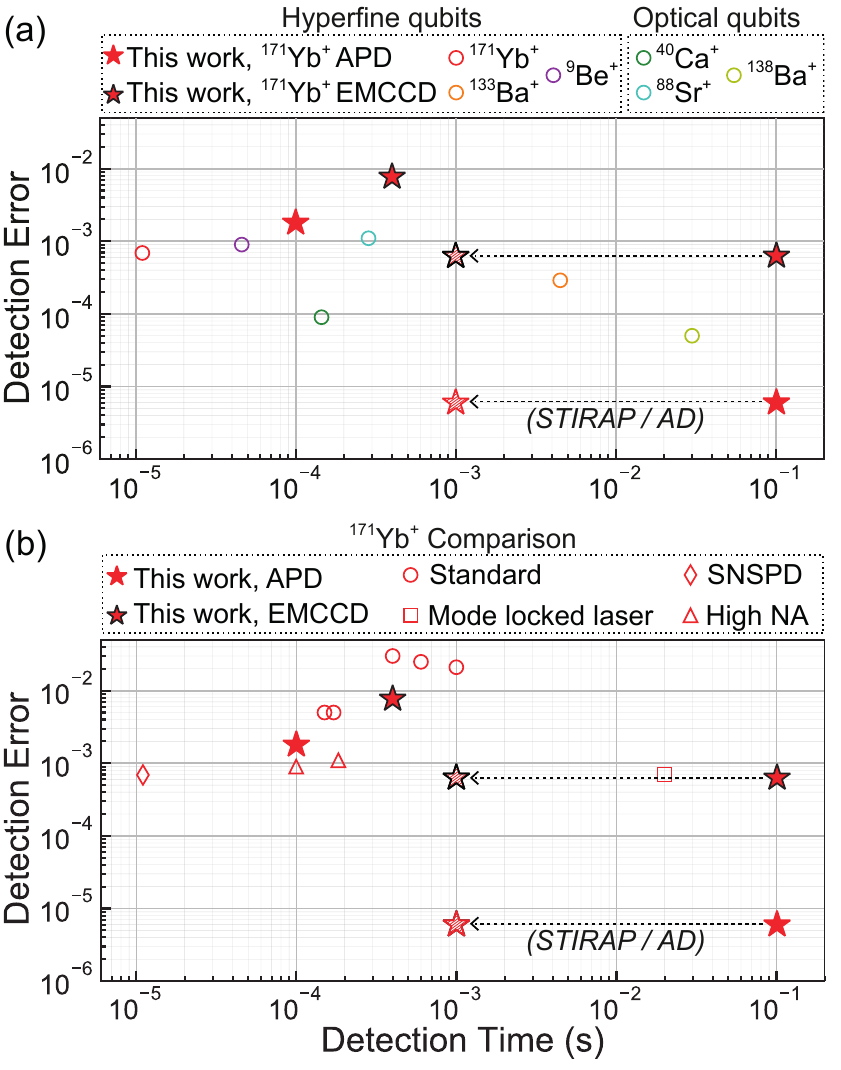}
    \caption{Comparison of single-qubit detection errors and times for trapped-ion qubits. (a) Overview of representative results for different hyperfine qubits, with $\Yb$~\cite{Crain:2019}, $\ion{Ba}{133}$~\cite{Christensen:2020} and $\ion{Be}{9}$~\cite{Todaro:2020}, and optical qubits, with $\ion{Ca}{40}$~\cite{Myerson:2008}, $\ion{Sr}{88}$~\cite{Keselman:2011} and $\ion{Ba}{138}$~\cite{Zhukas:2020}. Star markers represent the work presented here in $\Yb$ using the APD (red outline) and the EMCCD (black outline). The shaded star markers at 1~ms detection are achieved by detecting in the long-lived $\LS{F}{7/2}$ manifold in $\Yb$. Currently, this procedure requires $\sim100$~ms (filled star markers) for shelving, but additional laser hardware should reduce this time by two orders of magnitude using STIRAP or an active depopulation technique (AD) (see Sec.~\ref{sec:F-State-Shelving}).
    (b) Results for $\Yb$, with circle markers representing standard detection techniques that can be straightforwardly implemented without advanced detector technology~\cite{Olmschenk:2007, Ejtemaee:2010, Wolk:2015, Seif.2018, Ding.2019}. The work presented here, shown as star markers, similarly requires no additional detector hardware. 
    The remaining markers require special hardware such as a mode-locked laser~\cite{Roman:2020} (square marker), superconducting nanowire single photon detectors (SNSPDs)~\cite{Crain:2019} (diamond marker), or a high NA objective~\cite{Noek.2013} (triangle markers).}
    \label{fig:detection_fidelities}
\end{figure}

In $\Yb$ the measurement-fidelity limiting factor is leakage between the hyperfine qubit levels during detection. The two logical states encoded in the $\LS{S}{1/2}$ levels can be distinguished by state-selective fluorescence induced by a 369~nm laser resonant with the $\LS{S}{1/2}\ket{F=1} \leftrightarrow \LS{P}{1/2}\ket{F=0}$ transition, which, aside from a small branching ratio to $\LS{D}{3/2}$, forms a closed cycling transition (Fig.~\ref{fig:full_energy_levels}). However, the small hyperfine splitting of 2.11~GHz between adjacent $\LS{P}{1/2}$ levels results in a comparatively large off-resonant scattering probability causing leakage primarily from the bright state $\LS{S}{1/2}\ket{F=1}$ to the dark state $\LS{S}{1/2}\ket{F=0}$. The inverse occurs as well, but with a lower probability due to the effective $14.75$~GHz detuning. The dynamics of this asymmetric leakage during state detection in hyperfine qubits have been analyzed theoretically by Acton et al.~\cite{Acton.2005} and, more specifically for the case of $\Yb$, by W\"{o}lk et al.~\cite{Wolk:2015}. A further challenge in $\Yb$ is its low fluorescence yield compared to other isotopes without nuclear spin, such as $\ion{Yb}{174}$. Fluorescence increases with the strength of an applied magnetic field~\cite{Ejtemaee:2010}, but so does the magnetic field sensitivity of the clock transition encoding the qubit, negatively impacting the available phase coherence time ($T_2$).

The efficiency of a detection protocol can be quantified in two dimensions, through the measured detection error and the required detection time -- both of which should be ideally minimized for practical use in quantum computing. In Fig.~\ref{fig:detection_fidelities}(a) we show an overview of results reported in the trapped-ion field for both hyperfine and optical qubits together with the results (red stars) described in this manuscript using the APD (red outline) and the EMCCD (black outline). Figure~\ref{fig:detection_fidelities}(b) compares detection errors for the specific case of $\Yb$. Open-circle markers represent measurements achieved without special hardware for photon collection. The open-triangle markers show $\Yb$ detection fidelities achieved using a high NA objective~\cite{Noek.2013} and the open-square marker is a measurement in $\Yb$ using a mode-locked laser to achieve near background free detection in the dark periods between ultra-short pulses~\cite{Roman:2020}. The open-diamond markers use SNSPDs to collect photons from $\Yb$~\cite{Crain:2019} and $\ion{Be}{9}$~\cite{Todaro:2020}.  The work we report here exceeds the results achieved with ``standard'' detection hardware in $\Yb$, and if combined with high-speed optical pumping via STIRAP provides a route for field-leading performance in detection error at practically relevant measurement times.

\section{Experimental Setup}

\begin{figure*}[t!]
    \centering
    \includegraphics[scale=1]{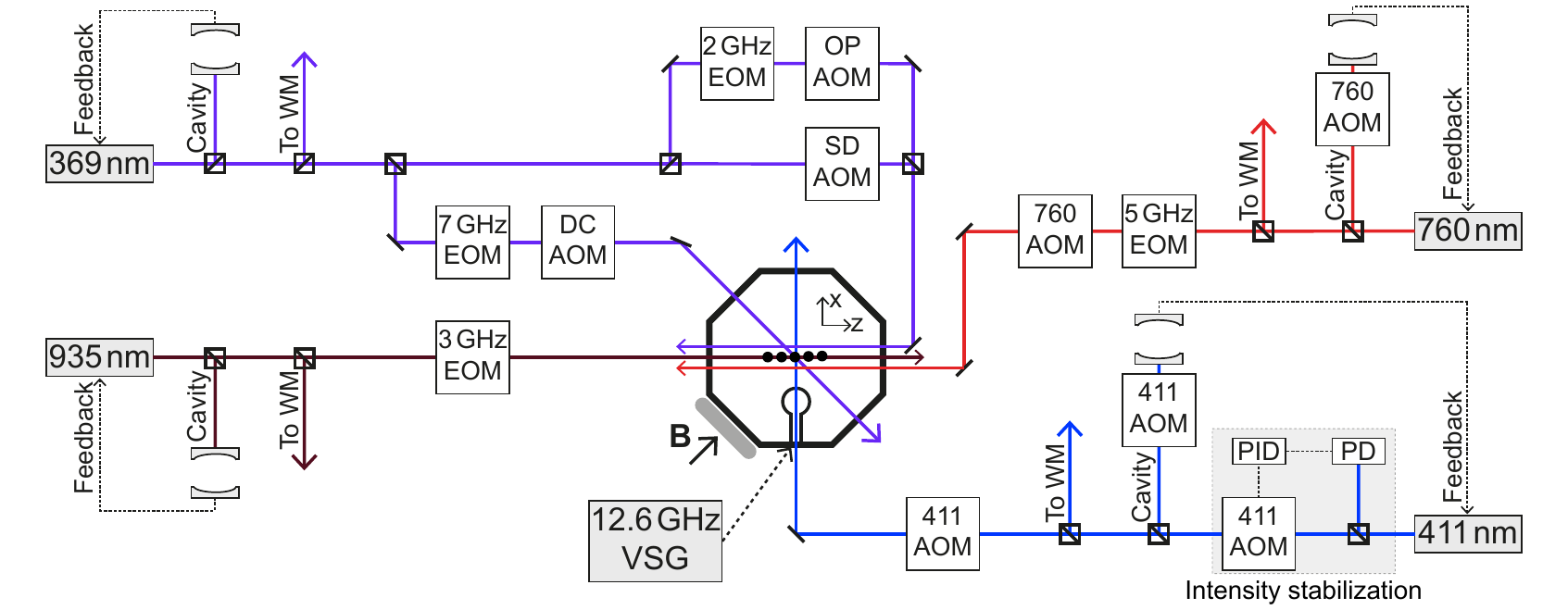}
    \caption{Simplified overview of the laser setup and trap geometry. Beam paths for the four relevant lasers are shown, including any pertinent EOMs and  AOMs, optical cavities, and the wavemeter (WM) pickoffs. The 369~nm laser is split into three beam paths for Doppler cooling (DC), optical pumping (OP), and state detection (SD). The 411~nm laser is intensity stabilized using a photodiode (PD) and a PID controller. 
    In the center, a schematic of the vacuum chamber containing the trap is shown with five ions along the trap axis. The bottom loop in the chamber is a 12.64~GHz microwave antenna driven by a vector signal generator (VSG). The grey bar at the lower left indicates a permanent magnet creating the quantization magnetic field.}
    \label{fig:experimental_setup}
\end{figure*}

We realize a qubit in the hyperfine clock states of $\Yb$, designating $\ket{0} \equiv \LS{S}{1/2}\ket{F=0, m_F=0}$ and $\ket{1} \equiv \LS{S}{1/2}\ket{F=1, m_F=0}$. Our experiments are carried out in a blade-style linear Paul trap with secular trap frequencies of \mbox{$\omega_{(x,y,z)}/2\pi = (1.6, 1.5, 0.5)$~MHz}. The magnetic quantization field is produced by a permanent magnet, creating a 440.9(6)~$\upmu$T field at the ion position, which was measured using the $\LS{S}{1/2}$ linear Zeeman shift of 13.98(1)~kHz/$\upmu$T~\cite{Meggers:1967}. Single-qubit operations are driven with a microwave field produced by a vector signal generator\footnote{Keysight E8267D} that is delivered through an in-vacuum loop antenna. Photons emitted during laser cooling and state detection are collected by a custom-made objective with an effective \mbox{NA = 0.56} and imaged onto either an APD\footnote{Laser Components COUNT-10B} or an EMCCD\footnote{Andor iXon Ultra 897}.

A simplified energy-level diagram showing the states and transitions in $\Yb$ relevant to this work is shown in Fig.~\ref{fig:full_energy_levels}, and a simplified schematic of the experimental setup is given in Fig.~\ref{fig:experimental_setup}. Doppler cooling (DC), optical pumping (OP), and state detection (SD) utilize the $\LS{S}{1/2} \leftrightarrow \LS{P}{1/2}$ transition; this requires a diode laser near 369.5~nm, nominally tuned to the inner \mbox{$\ket{F=1}\leftrightarrow\ket{F=0}$} transition that is split into three different beamlines. A second-order sideband from a 7.374~GHz electro-optic modulator (EOM) simultaneously excites the outer \mbox{$\ket{F=0}\leftrightarrow\ket{F=1}$} transition during Doppler cooling ensuring all manifold states are addressed. At the start of each experiment, following Doppler cooling, the ion is prepared in the qubit state $\ket{0}$ by adding a 2.105~GHz sideband via a separate EOM to the 369~nm laser; this optically pumps any population in $\ket{1}$ to $\ket{0}$ via \mbox{$\LS{P}{1/2}\ket{F=1}$}~\cite{Olmschenk:2007}. To directly measure the final qubit state, the 369.5~nm laser light is applied to the ion without any additional modulation to selectively excite $\ket{1}$, the ``bright'' qubit state. Occasional decays from $\LS{P}{1/2}$ to $\LS{D}{3/2}$ (0.5\%) remove the ion from the cooling cycle and logical state space, necessitating a repump laser at 935~nm  that is operated continuously. An EOM running at 3.067~GHz adds sidebands to the 935~nm laser to ensure both hyperfine levels of the $\LS{D}{3/2}$ are repumped. 

In this work, we introduce two additional lasers for the purpose of state detection: a 411~nm laser\footnote{Moglabs LDL-202 with a Fast Servo Controller (FSC)} for electron shelving from $\LS{S}{1/2}$ to $\LS{D}{5/2}$~\cite{Roberts:1999, Keller:2019, Feldker:2020, Roman:2020, Tan:2020}, and a 760~nm laser~\cite{Huntemann:2012, Jaua:2015, Mulholland:2019} for repumping from the long-lived $\LS{F}{7/2}$ state (\mbox{$\tau \approx 5.4$ years}~\cite{Roberts:2000}) via $\LSB{1}{D}{3/2}{3/2}$. The 760~nm laser replaces a 638~nm laser~\cite{Sugiyama:1999} commonly used for this purpose and gives the benefit of substantially faster repumping. An EOM driven at 5.258~GHz adds sidebands to the 760~nm laser to excite both $\LS{F}{7/2}$ hyperfine states. We stabilize both laser frequencies through Pound-Drever-Hall locks to cylindrical Fabry-P\'erot cavities\footnote{Stable Laser Systems, Boulder CO, USA} with a free spectral range of 1.5~GHz; the 411~nm (760~nm) cavity has a finesse of approximately 32,000 (1,000$-$2,000) and a drift rate of $\sim320$~mHz/s ($\sim3.2$~Hz/s). The ultra-low expansion spacer of the 411~nm cavity is temperature stabilized close to the minimum of its coefficient of thermal expansion (CTE) located at 38.2$^\circ$C. 
Absolute frequency measurements use a HeNe-calibrated wavemeter\footnote{HighFinesse WSU-10} with a 500~kHz precision, 10~MHz absolute accuracy at 760~nm and 177~MHz absolute accuracy at 411~nm (due to operating $>200$~nm from the 633~nm calibration wavelength).

\section{Atomic spectroscopy}

\label{sec:atomic_physics}

In order to implement electron-shelved detection using $\LS{D}{5/2}$ or $\LS{F}{7/2}$, we make use of a precision characterization of the $\LS{S}{1/2}\leftrightarrow\LS{D}{5/2}$ transition at 411~nm, which we report separately in a jointly submitted manuscript. 
Table~\ref{tab:atomic_physics} contains all of the relevant parameters that have been measured by our team, and we direct interested readers to Ref.~\cite{Tan:2020} for full details on the measurements of the 411~nm transition, including the lifetime, branching ratios, quadratic Zeeman coefficient and hyperfine constant of the $\LS{D}{5/2}$ states. From here on, we use simplified notation for the hyperfine levels in Dirac notation by omitting the $F$ and $m_F$ labels, shortening the labels to $\ket{F, m_F}$.

\renewcommand{\arraystretch}{1.5}
\begin{table}[t]
\begin{tabularx}{\columnwidth}{ 
   >{\raggedright\hsize=.67\columnwidth}X
   >{\raggedleft\arraybackslash}X}
     \hhline{==}
     Parameter & This work (exp.)
     \\ \hline
     411~nm frequency for \mbox{$\LS{S}{1/2}\ket{0,0}\leftrightarrow\LS{D}{5/2}\ket{2,0}$~(THz)}
     & 729.487752(177)
     \\
     411~nm frequency for \mbox{$\LS{S}{1/2}\ket{1,0}\leftrightarrow\LS{D}{5/2}\ket{3,0}$~(THz)}
     &  729.474917(177)
     \\
     Hyperfine constant of $\LS{D}{5/2}$~(MHz)
     &  -63.368(1)
     \\ \hline
     Linear Zeeman coefficient of $\LS{D}{5/2}\ket{3}$~(kHz/$\upmu\textrm{T}$)
     & 13.96(2) 
     \\
     Linear Zeeman coefficient of $\LS{D}{5/2}\ket{2}$~(kHz/$\upmu\textrm{T}$)
     & 19.61(3) 
     \\
     Quadratic  Zeeman  coefficient for $\LS{D}{5/2}\ket{3, 0}$~(Hz/$\upmu\textrm{T}^2$)
     & -0.350(1)
     \\ \hline
     Lifetime of $\LS{D}{5/2}\ket{3}$ (ms)
     & 7.1(4)
     \\ 
     Decay from $\LS{D}{5/2}\ket{3}$ to $\LS{S}{1/2}\ket{1}$ 
     & 17.6(4)\%
     \\
     Decay from $\LS{D}{5/2}\ket{3}$ to $\LS{F}{7/2}$ 
     & 82.4(4)\%
     \\
    Lifetime of $\LS{D}{5/2}\ket{2}$ (ms)
     & 7.4(4)
    \\
     Decay from $\LS{D}{5/2}\ket{2}$ to $\LS{S}{1/2}\ket{0}$ 
     & 11.1(3)\%
     \\
     Decay from $\LS{D}{5/2}\ket{2}$ to $\LS{S}{1/2}\ket{1}$ 
     & 7.4(3)\%
     \\
     Decay from $\LS{D}{5/2}\ket{2}$ to $\LS{F}{7/2}$ 
     & 81.6(4)\%
     \\
     \hline
     760~nm repumper center frequency (THz) after preparing $\LS{D}{5/2} \ket{3, 0}$
     & 394.430203(16)
     \\ 
     760~nm repumper center frequency (THz) after preparing $\LS{D}{5/2} \ket{2, 0}$
     & 394.424943(20)
     \\ \hhline{==}
\end{tabularx}
\caption{Relevant parameters for electron-shelved state detection via the $\LS{S}{1/2}\leftrightarrow\LS{D}{5/2}$ transition in $\Yb$.
}
\label{tab:atomic_physics}
\end{table}

\begin{figure}[h]
    \centering
    \includegraphics[scale=1]{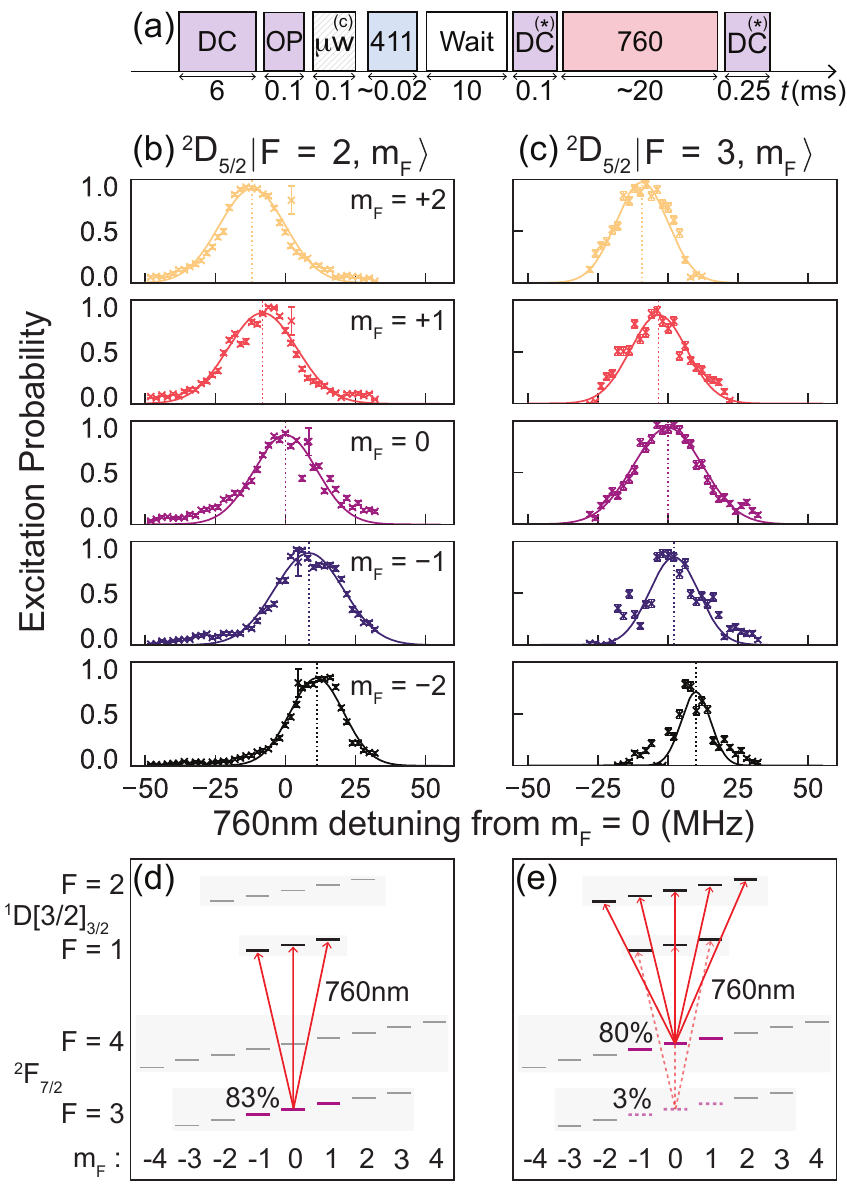}
    \caption{Measurement of the 760~nm $\LS{F}{7/2}$ repump transition. (a) Schematic for the measurement protocol using the 760~nm, 411~nm, as well as Doppler cooling (DC) and optical pumping (OP) light at 369~nm. High power, on resonance Doppler cooling light is utilized for detection (DC$^{(*)}$). A microwave $\pi$ pulse ($\upmu$w) is used in (c) to prepare $\LS{S}{1/2}\ket{1, 0}$. \mbox{(b, c)} Transitions observed at 760~nm after shelving the ion to \mbox{$\LS{D}{5/2}\ket{2, m_F}$} or \mbox{$\LS{D}{5/2}\ket{3, m_F}$}. For both transitions, the five Zeeman states $m_F=0, \pm1, \pm2$ are individually prepared by a 411~nm pulse and the recovery probability for a given 760~nm laser frequency is shown with error bars are derived from quantum projection noise. The center frequency of each excitation pathway from $\LS{F}{7/2}$ is extracted from a Gaussian fit. \mbox{(d, e)} Energy diagrams illustrating the possible $\LS{F}{7/2}$ states to which the ion can decay from $\LS{D}{5/2}\ket{F, m_F=0}$ as well as the allowed 760~nm excitations from $\LS{F}{7/2}\ket{F, m_F=0}$.}
    \label{fig:760_spectra}
\end{figure}

After each experiment involving the $\LS{D}{5/2}$ state, any population that has decayed to $\LS{F}{7/2}$ must be returned to the qubit manifold. There are several possible transitions over a range of wavelengths that can be driven to achieve this goal: 638~nm~\cite{Gill:1995}, 760~nm~\cite{Huntemann:2012},  828~nm~\cite{Sugiyama:1999}, or 864~nm~\cite{Taylor:1997}. Here, we employ a 760~nm laser as it has been observed to have the most rapid clear-out time. This phenomenology owes to its excited energy level $\LSB{1}{D}{3/2}{3/2}$ exhibiting a short upper-state lifetime (29~ns~\cite{Berends:1993}), and a decay path primarily to the $\LS{S}{1/2}$ ground states rather than other D-levels~\cite{Sugiyama:1999}.

In Fig.~\ref{fig:760_spectra}, we present the measured spectra for the 760~nm $\LS{F}{7/2} \leftrightarrow \LSB{1}{D}{3/2}{3/2}$ transition; the measurement protocol is illustrated schematically in panel (a). We first prepare the ion in one of the $\LS{D}{5/2}$ levels using the 411~nm shelving laser. A subsequent 10~ms wait period allows the ion to decay to either $\LS{S}{1/2}$ or $\LS{F}{7/2}$. Following the wait period, a 100~$\upmu$s period of high power, on resonance Doppler cooling induces fluorescence in cases where the ion has not decayed to $\LS{F}{7/2}$, and we discard these experiments in post-processing. The 760~nm laser is then applied to clear out the $\LS{F}{7/2}$ state at a frequency adjusted through AOMs. To detect the final state, high power, on resonance Doppler cooling light is used to determine whether the ion has returned to the $\LS{S}{1/2}$ manifold or remains in $\LS{F}{7/2}$. The probability of excitation from $\LS{F}{7/2}$ is plotted in Fig.~\ref{fig:760_spectra}(b),(c) for states prepared in different $m_F$ levels of 
(b) $\LS{D}{5/2} \ket{2}$ and (c) $\LS{D}{5/2} \ket{3}$. 
The energy level diagrams in Fig.~\ref{fig:760_spectra}(d),(e) show the potentially occupied $\LS{F}{7/2}$ states populated through the dipole decay from (d) $\LS{D}{5/2}\ket{2, 0}$ or (e) $\LS{D}{5/2}\ket{3, 0}$ with the corresponding decay probabilities from the $\LS{D}{5/2}$ levels~\cite{Roberts:1999, Tan:2020}, and the possible repump pathways from $\LS{F}{7/2}$ $\ket{F, m_F=0}$. As $\LS{F}{7/2}\leftrightarrow \LSB{1}{D}{3/2}{3/2}$ is an electric quadrupole (E2) coupling $\Delta m_F= 0, \pm1, \pm2$ transitions are possible. Consequently, given a prepared $m_F$ level in $\LS{D}{5/2}$, the 760~nm peaks shown in the measurements comprise between 6 and 22 possible unresolved transitions of varying probabilities. 
The center frequencies of the transitions, obtained from Gaussian fits, are reported in Tab.~\ref{tab:atomic_physics} with a $\sim25\times$ improvement in precision relative to previous results~\cite{Mulholland:2019}.

\section{\texorpdfstring{$\Yb$}{171Yb+} state detection using electron shelving}
\label{sec:qubit_SD}

The standard detection protocol used to discriminate between the qubit states $\ket{0}$ and $\ket{1}$ in $\Yb$ relies on detecting state-selective laser-induced fluorescence at 369~nm. Ideally, the collected photons result in two well-separated Poisson distributions corresponding to the different states. However, leakage between the qubit levels due to off-resonant excitation during detection create one-sided tails on the photon distributions that overlap and thereby lead to detection errors. 

In the following, we augment the standard detection method by prepending the measurement with pulses at 411~nm in order to transfer population from one qubit state to a metastable level. State detection is then performed using 369~nm light that drives all transitions between the $\LS{S}{1/2}$ and $\LS{P}{1/2}$ manifolds, effectively eliminating off-resonant excitations. Here, the achievable detection fidelity is limited by the shelving transfer accuracy and the finite lifetime of the metastable state. As indicated in Tab.~\ref{tab:atomic_physics}, the $\LS{D}{5/2}$ state lifetime is $\sim7$~ms, after which it decays to $\LS{F}{7/2}$ ($\sim82\%$) or $\LS{S}{1/2}$ ($\sim18\%$). This presents two possibilities for an electron-shelving based detection protocol: (1) transfer the population of one qubit state to $\LS{D}{5/2}$ and detect for duration $t\ll 7$~ms before any significant decay occurs; 
or (2) optically pump the $\ket{1}$ state to $\LS{F}{7/2}$ via $\LS{D}{5/2}$. In this section, we describe the detection protocols using both shelving methods, discuss different software-based techniques to improve state discrimination under these protocols, and finally compare the achieved state detection fidelities.

\subsection{Electron-shelved detection in \texorpdfstring{$\LS{D}{5/2}$}{D5/2}}
\label{sec:d-state_detection}

\begin{figure}[t]
    \centering
    \includegraphics[scale=1]{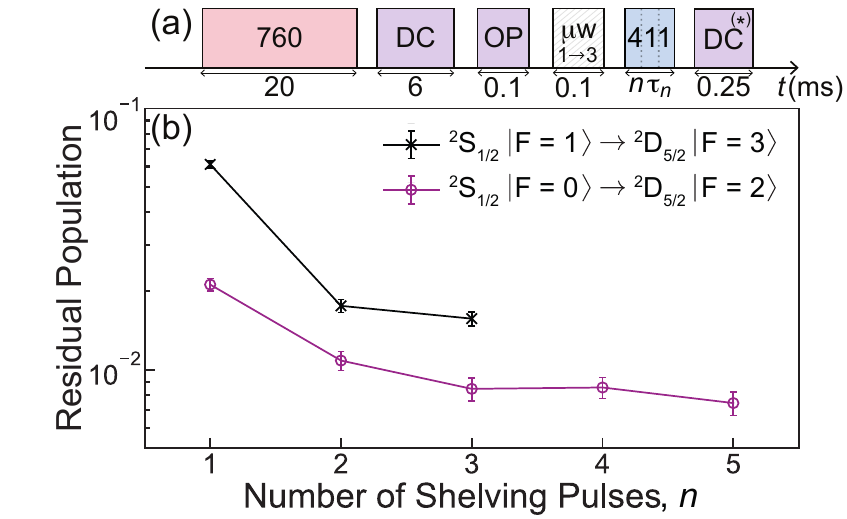}
    \caption{Shelving efficiency of multiple $\pi$ pulses. (a) Schematic showing the experimental sequence including $n$ 411~nm shelving $\pi$ pulses of length $\tau_{n}$, a microwave pulse ($\upmu$w) used to initialize the ion in $\LS{S}{1/2}\ket{1, 0}$ before the $\LS{S}{1/2}\ket{1}\rightarrow\LS{D}{5/2}\ket{3}$ transition, and 369~nm light used for Doppler cooling (DC), optical pumping (OP) and detection (DC$^{(*)}$). (b) The residual population in the qubit manifold $\LS{S}{1/2}$ after shelving via (black) $\LS{S}{1/2}\ket{1}\rightarrow\LS{D}{5/2}\ket{3}$ using three successive shelving pulses, $\Delta m_F=0, \pm2$, and (pink) $\LS{S}{1/2}\ket{0}\rightarrow\LS{D}{5/2}\ket{2}$ using five successive shelving pulses, $\Delta m_F=0, \pm1, \pm2$. The maximum shelving fidelity is 99.3(1)\% after five shelving pulses to $\LS{D}{5/2}\ket{2}$. Error bars are calculated from quantum projection noise.}
    \label{fig:shelving_success}
\end{figure}
The first protocol requires us to achieve an effective transfer of population from either hyperfine level in the qubit manifold to the metastable $\LS{D}{5/2}$ state. We begin by measuring the effectiveness of applying a $\pi$ pulse on two transitions:  $\LS{S}{1/2}\ket{1}\rightarrow\LS{D}{5/2}\ket{3}$ and $\LS{S}{1/2}\ket{0}\rightarrow\LS{D}{5/2}\ket{2}$ (Fig.~\ref{fig:shelving_success}(a)). The linewidth of the 411~nm laser as well as the non-zero temperature of the motional modes limit the shelving efficiency associated with a single $\pi$ pulse. Thus, in order to maximize population transfer, we implement a series of $\pi$ pulses tuned to address multiple $\LS{D}{5/2}$ Zeeman levels. When shelving via $\LS{S}{1/2}\ket{0} \rightarrow \LS{D}{5/2}\ket{2}$, five Zeeman transitions can be driven successively, $\Delta m_F = 0, \pm1, \pm2$. By contrast, the $\LS{S}{1/2}\ket{1}\rightarrow\LS{D}{5/2}\ket{3}$ transition only allows three successive pulses on $\Delta m_F = 0, \pm2$ to be used, as the first-order Zeeman shift is approximately equal for the upper and lower states ($\sim14$~kHz/$\upmu$T). If the $\Delta m_F = \pm1$ transitions are excited, then any population initially transferred  to $\LS{D}{5/2}\ket{3, 0}$ will be \textit{de-shelved} by the subsequent pulses.

The residual population in the qubit manifold after shelving is plotted in Fig.~\ref{fig:shelving_success}(b). As the number of shelving pulses is increased, the shelving fidelity improves from 93.9(2)\% to 98.4(1)\% for $\LS{S}{1/2}\ket{1} \rightarrow \LS{D}{5/2}\ket{3}$ (black), and from 97.9(2)\% to 99.3(1)\% for $\LS{S}{1/2}\ket{0} \rightarrow \LS{D}{5/2}\ket{2}$ (pink). 
The clock transition between the $m_F=0$ states is driven first as its first-order magnetic-field insensitivity allows for the highest state transfer probability. In general, we observe no significant difference when changing the order of the subsequent pulses tuned to other Zeeman levels.

When using this protocol in large multi-qubit registers, it is critical to ensure that the population transfer efficiency remains high for all ions throughout the experiment. Its effectiveness will be limited by laser frequency drifts and variations in coupling strength due to laser intensity or polarization gradients across the ion string. As both of these effects are often slowly varying or even static ``systematic'' errors, mitigation through advanced pulse sequences~\cite{Tycko.1983,Levitt.1986,Kabytayev.2014,Ball:2020} can be considered in addition to regular calibration. Another commonly used routine for accurate population transfer is Rapid Adiabatic Passage (RAP)~\cite{Vitanov.2001,Wunderlich:2007, Noel:2012,Fuerst:2020}. This technique involves linearly sweeping the frequency of the shelving laser, while simultaneously shaping the amplitude of pulse to follow a Gaussian profile. The procedure is more robust to systematic errors in the pulse frequency or length than a simple $\pi$ pulse, at the cost of enhanced sensitivity to high-frequency and dephasing errors~\cite{Lu:2013, Soare:2014, Edmunds:2020}. 

\begin{figure}
    \centering
    \includegraphics[scale=1]{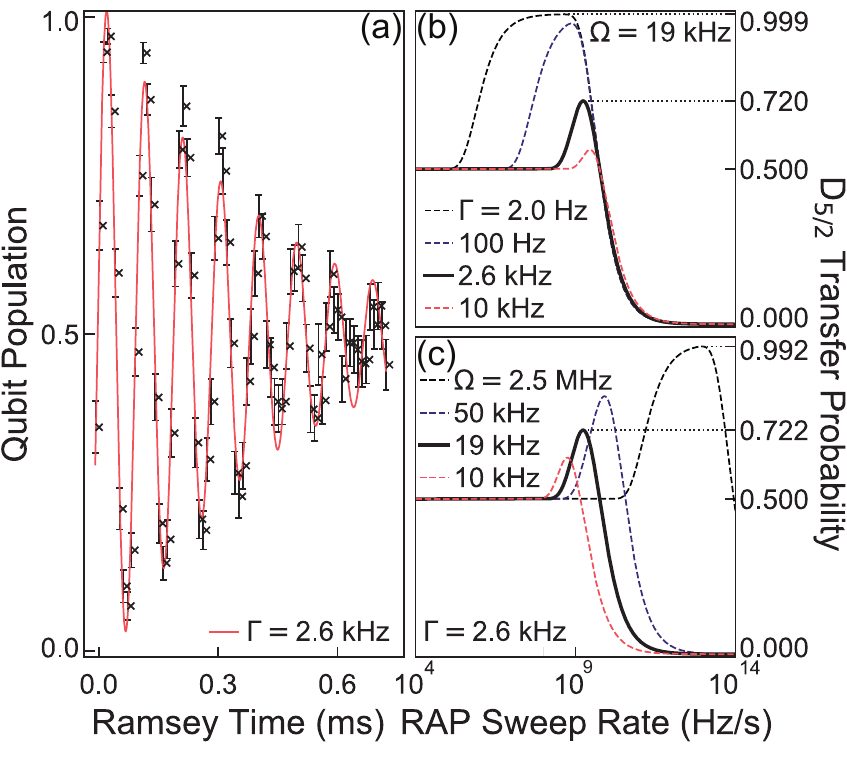}
    \caption{Transfer fidelity using Rapid Adiabatic Passage (RAP). (a) Ramsey interferometry using the 411~nm laser on the $\LS{S}{1/2}\ket{0,0}\leftrightarrow\LS{D}{5/2}\ket{2,0}$ transition, giving an inverse laser coherence $\Gamma=2.6$~kHz. (b, c) Maximum population transfer using RAP for (b) fixed Rabi frequency $\Omega=19$~kHz and varying $\Gamma$, or (c) fixed $\Gamma=2.6$~kHz and varying $\Omega$. In both figures, the solid black line is the result of our current experimental parameters and the black dashed line shows the maximum achievable transfer.}
    \label{fig:RAP_transfer}
\end{figure}

In Fig.~\ref{fig:RAP_transfer} we examine the theoretical maximum transfer fidelity using RAP for different inverse laser coherence times, $\Gamma$, and Rabi frequencies, $\Omega$. The probability of transfer using a Landau-Zener model~\cite{Zener:1932} for RAP is given by
\begin{equation}
    P_{LZ} = 1 - e^{-\pi^2\Omega^2/\alpha},
\end{equation}
where $\alpha$ is the frequency sweep rate used for the RAP pulse. To incorporate the effect of a finite laser linewidth, the theory is modified to include a Markovian noise bath in a two-level dephasing model~\cite{Lacour:2007, Noel:2012}. The transfer probability now depends on the inverse of the laser coherence, $\Gamma$, becoming
\begin{equation}
    \label{eq:RAP}
    P = \frac{1}{2} \left( 1-e^{-2\pi^2\Gamma\Omega/\alpha} \right) + e^{-2\pi^2\Gamma\Omega/\alpha} P_{LZ},
\end{equation}
which results in a sharp dropoff in transfer fidelity at lower sweep rates. 

In our experiment we record a Rabi frequency of 19~kHz on the $\LS{S}{1/2}\ket{0, 0}\rightarrow\LS{D}{5/2}\ket{2, 0}$ clock transition and a 0.392~ms phase coherence time. This number is inferred from Ramsey interferometry shown in Fig.~\ref{fig:RAP_transfer}(a) and corresponds to an inverse coherence time of $\Gamma = 2.6$~kHz, which we attribute to laser linewidth. In Fig.~\ref{fig:RAP_transfer}(b),(c) we plot the calculated transfer fidelity against the sweep rate, as given by Eqn.~\eqref{eq:RAP}. Given our parameters, $\Gamma=2.6$~kHz, $\Omega=19$~kHz, we could achieve a maximum transfer fidelity of 0.72 (solid black lines), which is significantly worse than the fidelity of a single $\pi$ pulse ($\sim$98\%). To achieve $>99$\% transfer fidelity, we would either need to improve our laser coherence to 2~Hz or increase our Rabi frequency to 2.5~MHz (dashed black lines); both of these are unfeasible in our current system. However, more reasonable parameter regimes achieving the same target can be found with two-dimensional parameter analysis, e.g. reducing the inverse coherence time to $\sim$130~Hz and increasing the Rabi frequency to 100~kHz. This would be achievable using higher laser lock-bandwidth and a different laser source, respectively. We further plan to investigate numerically optimized robust control waveforms~\cite{Ball:2020} to improve state transfer efficiency.

\subsection{Electron-shelved detection in \texorpdfstring{$\LS{F}{7/2}$}{F7/2}}
\label{sec:F-State-Shelving}
Another attractive option for electron-shelved detection in $\Yb$ uses the $\LS{F}{7/2}$ state with a lifetime in excess of 5~years~\cite{Roberts:2000}, enabling longer detection periods, while still eliminating off-resonant scattering. Given that $\LS{S}{1/2} \rightarrow \LS{F}{7/2}$ is an electric octupole transition, direct shelving to this state requires ultra-stable laser systems generally only available in specialized frequency metrology laboratories~\cite{Huntemann:2012,Huntemann.2012hb,Fuerst:2020}. 

To investigate this level for state detection without such a laser, we optically pump the $\ket{1}$ qubit state to the $\LS{F}{7/2}$ manifold via $\LS{D}{5/2}\ket{3}$ using 411~nm light. 
As before with the electron-shelved detection in $\LS{D}{5/2}$, once the population has been transferred to $\LS{F}{7/2}$, we use high power light resonant with the entire $\LS{S}{1/2}$ and $\LS{P}{1/2}$ manifolds to measure laser induced fluorescence from the population remaining in the qubit manifold.  

The incoherent shelving process used by us requires $\sim100$~ms to ensure $>$99.9\% population transfer to $\LS{F}{7/2}$, which makes it impractical for use in quantum computing. 
One way to achieve fast shelving to the F-state is via a STIRAP-like scheme~\cite{Vitanov.2017} to a state that rapidly decays to the $\LS{F}{7/2}$ manifold. To implement such a scheme, one could use a laser at $410$~nm connecting the $\LS{D}{3/2}$ metastable state to the $^{1}\mathrm{[5/2]_{5/2}}$ level~\cite{Sugiyama:1999,Schacht.2015} and combine it with light at 435~nm connecting the $\LS{S}{1/2}$ manifold to the $\LS{D}{3/2}$ state. As this scheme relies on the rapid decay of the $^{1}\mathrm{[5/2]}_{5/2}$ state, it could be executed repeatedly akin to optical pumping, ensuring a high transfer efficiency to the long lived $\LS{F}{7/2}$ state.
Alternatively, a pulsed two-stage scheme could be used, where multiple shelving pulses (similar to Fig.~\ref{fig:shelving_success}) are combined with active depopulation (AD in Fig.~\ref{fig:detection_fidelities}) of the $\LS{D}{5/2}$ state via resonant light at $3.4~\upmu\mathrm{m}$~\cite{Roberts:1999}. 

\subsection{State-detection protocol comparison}

In this section, we now compare the detection fidelity for three different detection protocols: (1) $\LS{S}{1/2}$ standard detection with light resonant only with the $\ket{1}$ state in the qubit manifold, (2) $\LS{D}{5/2}$-shelved detection with the Doppler cooling laser tuned on resonance at high power after shelving the $\ket{0}$ qubit state to $\LS{D}{5/2}\ket{2}$ via five successive $\pi$ pulses to different Zeeman states, and (3) $\LS{F}{7/2}$-shelved detection with the resonant Doppler cooling laser after incoherently shelving  the $\ket{1}$ qubit state via $\LS{D}{5/2}$. 

In all cases measurements are conducted using either an APD recording global fluorescence or an EMCCD camera providing the spatially resolved information required for experiments with multi-qubit registers. Except for protocol (3), we compare the performance of a simple threshold-based detection with a time-resolved maximum likelihood analysis for the APD data. For all three protocols, we compare thresholding and a classifier-based software routine for analysis of EMCCD data.

The detection error is calculated by interleaving preparation of a dark and bright qubit state and averaging the respective errors. We define the dark state error, $\epsilon_d$, as the fraction of points prepared in the dark state that are recorded as bright, and the bright state error $\epsilon_b$ accordingly. The overall detection error is then quantified as $\epsilon = (\epsilon_d+\epsilon_b)/2$. 
In order to derive a threshold value and train the image classifier, five percent of measured data are dedicated to calibration/training, with the analysis being conducted on the remaining 95\% of the data. 

\subsubsection{APD-based detection}

\begin{figure}[t]
    \centering
    \includegraphics[scale=1]{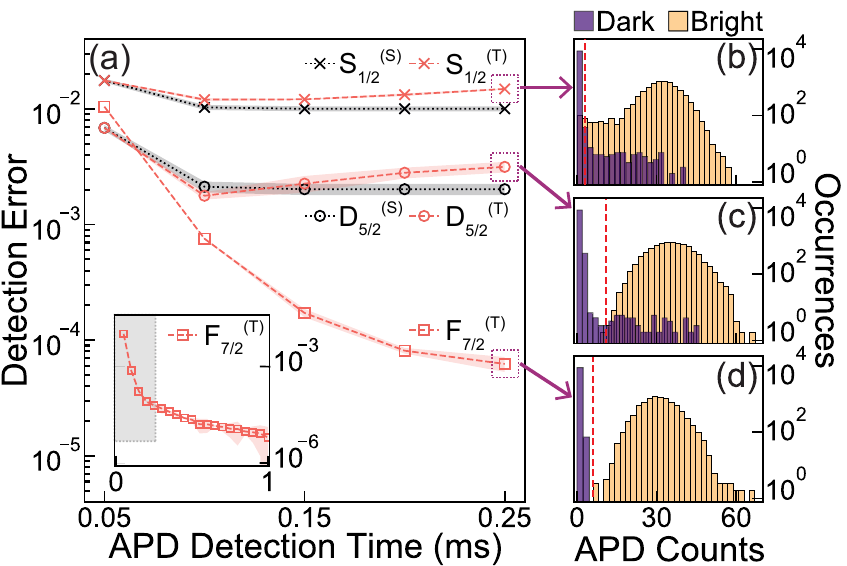}
    \caption{APD detection for three different detection protocols and two analysis methods. (a) Detection error as a function of detection time for $\LS{S}{1/2}$ standard detection with thresholding (red crosses), ${\textrm{S}_{1/2}}^\textrm{(T)}$, and with subbinning (black crosses), ${\textrm{S}_{1/2}}^\textrm{(S)}$; $\LS{D}{5/2}$-shelved detection with thresholding (red circles), ${\textrm{D}_{5/2}}^\textrm{(T)}$, and with subbinning (black circles),
    ${\textrm{D}_{5/2}}^\textrm{(S)}$; and $\LS{F}{7/2}$-shelved detection with thresholding only (red squares), ${\textrm{F}_{7/2}}^\textrm{(T)}$.  
    For both the bright and dark state measurements 20,000 points are taken when using $\LS{S}{1/2}$ and $\LS{D}{5/2}$ detection, while for $\LS{F}{7/2}$ we measure 1,000,000 points to resolve errors at the \num{1e-6} level. (Inset) Extended $\LS{F}{7/2}$-shelved results for detection periods up to 1~ms. The gray region indicates the size of the main panel.
    (b)-(d) Bright and dark state distributions with 250~$\upmu$s detection time for the three methods $\LS{S}{1/2}, \LS{D}{5/2}, \LS{F}{7/2}$ respectively. The red dashed line represents the optimal threshold between the photon distributions determined from five percent of the data. Error bands in (a) show the standard deviation resulting from 20 different optimization runs to find the optimal threshold by subsampling the data.}
    \label{fig:optimizing_APD_detection_time}
\end{figure}

To find the lowest detection error achievable with the APD, we vary the length of the detection period for the three different protocols as shown in Fig.~\ref{fig:optimizing_APD_detection_time}(a). 
In the simplest protocol (1) of state-dependent fluorescence detection in the qubit manifold, the photon count histograms of the bright (dark) state are described by a Poissonian distribution modified with a tail towards the dark (bright) state counts~\cite{Acton.2005}. This leads to a significant overlap visible in Fig.~\ref{fig:optimizing_APD_detection_time}(b) and correspondingly a large detection error.
When one of the qubit states is shelved by five successive $\pi$ pulses to $\LS{D}{5/2}\ket{2}$ in protocol (2), the application of detection light resonant with the entire $\LS{S}{1/2}$ manifold eliminates the off-resonant scattering and hence the decay tail of the bright state completely, which is shown in Fig.~\ref{fig:optimizing_APD_detection_time}(c). A residual but suppressed tail from the dark-state distribution remains due to decays from $\LS{D}{5/2}$ back to the qubit manifold, with an 18\% branching ratio.
Lastly, incoherent shelving to the long-lived $\LS{F}{7/2}$ state in protocol (3) maximally suppresses the decay tails on both distributions as illustrated in Fig.~\ref{fig:optimizing_APD_detection_time}(d).  To reduce the prohibitively long shelving times to $\LS{F}{7/2}$ under optical pumping via $\LS{D}{5/2}$ in these measurements, the ion is shelved only once every 1000 points before the dark state error is measured. The ion is then returned to the qubit manifold with the 760~nm laser in order to measure the bright state error for another 1000 points. These interleaved blocks are repeated 1000 times yielding a total of $10^6$ datapoints for each case. The total detection period is extended up to 1~ms for $\LS{F}{7/2}$-shelved detection as illustrated in the inset of Fig.~\ref{fig:optimizing_APD_detection_time} (a).

To further improve on the detection error in the two protocols that show state decays, we also implement a maximum likelihood estimation based on time-resolved data~\cite{Wolk:2015}, referred to hereafter as ``subbinning''. Here, additional information about the decay dynamics during a measurement period is obtained by dividing the $250~\upmu$s overall detection period into five smaller ``subbins'' of length $50~\upmu$s. This approach improves the ability to identify decay dynamics and allows for better discrimination of dark counts originating from electronic noise or cosmic particles. For the analysis, we further require an independent measurement of average count rates and the off-resonant scattering rates for the bright and dark states. Given our standard parameters for standard $\LS{S}{1/2}$ detection, we measure decay times from the bright and dark states of $\tau_B \approx 2$~ms and $\tau_D \approx 30$~ms, respectively. 
For $\LS{D}{5/2}$-shelved detection, $\tau_D$ is given by the $\sim 7$~ms upper state lifetime of the shelved state, while $\tau_B$ has an effectively infinite value. Given that there is no measurable decay during detection after shelving to $\LS{F}{7/2}$, we do not perform time-resolved analysis under that protocol.

\subsubsection{EMCCD-based detection}

To obtain spatially resolved measurements as required for multi-qubit experiments, we employ an EMCCD detector. Camera-based detection requires the identification of regions of interest (ROI) for pixel-based analyses. We locate these through Gaussian fits to the 2D ion location(s) of calibration images of (a) bright ion(s). Further processing then happens only on ROI data extracted from the full camera images, which decreases processing time and can readily be parallelized. 

\begin{figure}[t] 
    \centering
    \includegraphics[scale=1]{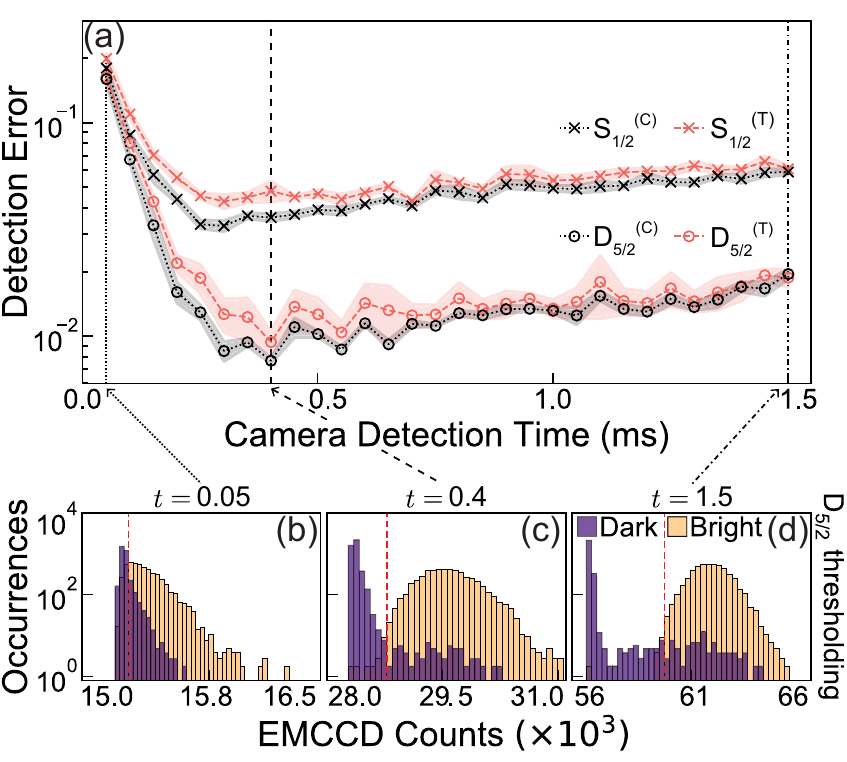}
    \caption{EMCCD based detection for two different detection protocols and analysis methods. (a) Detection error using $\LS{S}{1/2}$ standard detection with thresholding (red crosses), ${\textrm{S}_{1/2}}^\textrm{(T)}$, and the image classifier (black crosses), ${\textrm{S}_{1/2}}^\textrm{(C)}$; and electron-shelved detection after quintuple shelving to $\LS{D}{5/2}\ket{2}$ with thresholding (red circles), ${\textrm{D}_{5/2}}^\textrm{(T)}$, and the classifier, ${\textrm{D}_{5/2}}^\textrm{(C)}$. The lines show the means after sampling five different sets of training data, and shaded bands are $\pm 1\sigma$.  5000 measurements are taken at each point for both the bright and dark prepared states.  Five percent of the total data set is employed for training and used for identification of hot pixels, thresholds, and classifier training.  (b)-(d) Camera histograms for $\LS{D}{5/2}$-shelved detection at three detection times, illustrating how an optimum time is found for thresholding. For each histogram, the dashed red line marks the optimal threshold between ``dark'' and ``bright''.}
    \label{fig:camera_detect_time}
\end{figure}

As an ROI consists of multiple pixels, the thresholding method integrates their values over a certain number of ``hot'' pixels (corresponding to a subset of the brightest pixels) to obtain a measure of total counts in a given ROI. Alternatively, a set of calibration images obtained using ions prepared in the dark and bright states can be used to train a Random Forest classifier~\cite{scikit-learn} for each ROI in order to identify $\ket{0}$ and $\ket{1}$. If trained on reliable data, this method is expected to be superior to the simple thresholding model, as it will consider not just the net fluorescence in the region of interest but also $\textit{correlations}$ between the counts on different pixels. 

We compare the two spatial analysis methods across the three different detection protocols:  standard $\LS{S}{1/2}$, $\LS{D}{5/2}$-shelved and $\LS{F}{7/2}$-shelved. In Fig.~\ref{fig:camera_detect_time}(a), we evaluate the measured detection error as a function of detection time for the first two protocols.  The $\LS{S}{1/2}$ standard detection (cross markers) is compared to the $\LS{D}{5/2}$-shelved detection (open circles) and each dataset is analyzed using the thresholding method (red) and the image classifier (black).  These data clearly show that the shelved detection is superior to standard and that the image classifier can yield appreciable improvements under short to intermediate detection times. This is likely related to the a larger number of mislabeled training images due to state decays, potentially offering room for further improvement.

The change in detection error at different detection times can be understood by examining the bright and dark histograms for $\LS{D}{5/2}$-shelved detection (Fig.~\ref{fig:camera_detect_time}(b)-(d)). At short detection times, the distributions have a large overlap, with electrical noise in the camera dominating the signal (Fig.~\ref{fig:camera_detect_time}(b), 0.05~ms detection period). At long detection periods (Fig.~\ref{fig:camera_detect_time}(d), 1.5~ms), the bright distribution mean has increased sufficiently to separate it from the dark distribution, but state decays become dominant due to the 7~ms lifetime of $\LS{D}{5/2}$ producing a decay tail from the dark distribution.  At the optimum detection period, these two error contributions are balanced (Fig.~\ref{fig:camera_detect_time}(c), 0.4~ms detection period).

\begin{table*}[t]
    \begin{tabularx}{\textwidth}{c *{4}{>{\centering\arraybackslash}X c}}
    \hhline{=========}
         \multirow{2}{*}{\backslashbox{Detection}{Analysis}}
         & \multicolumn{2}{c}{\qquad EMCCD thresholding}
         & \multicolumn{2}{c}{\qquad EMCCD classifier}
         & \multicolumn{2}{c}{\qquad APD thresholding}
         & \multicolumn{2}{c}{\qquad APD subbinning}
         \\
         & Error
         & $t$~(ms)
         & Error
         & $t$~(ms)
         & Error
         & $t$~(ms)
         & Error
         & $t$~(ms)
         \\ \hline
         $\LS{S}{1/2}$ standard 
         & \num{4.3(3)e-2}  & 0.3 
         & \num{3.3(2)e-2}  & 0.3 
         & \num{1.20(6)e-2} & 0.1 
         & \num{1.00(5)e-2} & 0.15
         \\ 
         $\LS{D}{5/2}$-shelved 
         & \num{9(1)e-3}   & 0.4 
         & \num{7.7(2)e-3} & 0.4 
         & \num{1.8(2)e-3} & 0.1 
         & \num{2.0(2)e-3} & $\geq$0.15
         \\ 
         $\LS{F}{7/2}$-shelved 
         & \num{2(1)e-3} & 1 
         & \num{6.3(3)e-4} & 1 
         & \num{6(7)e-6} & 1 
         & - & -
         \\ \hhline{=========}
    \end{tabularx}
    \caption{State detection errors and optimum detection times on the EMCCD and APD using three detection protocols: (1) standard detection in $\LS{S}{1/2}$, (2) electron-shelved detection in $\LS{D}{5/2}$, and (3) incoherently shelved detection in $\LS{F}{7/2}$. These are compared for different analysis methods: basic thresholding, camera image classification, and time-resolved subbinning.}
    \label{tab:detection_errors}
\end{table*}

\subsection{Summary of results}

We summarize our findings of the lowest measured errors for all detection protocols and analysis methods in Table~\ref{tab:detection_errors}. 
Overall, we achieve a 5.6$\times$ improvement in fidelity using electron-shelved detection in $\LS{D}{5/2}$ compared to standard $\LS{S}{1/2}$ detection, measuring an error of \num{1.8(2)e-3} with a 100~$\upmu$s detection time on the APD. When using the EMCCD, we record a minimum error of \num{7.7(2)e-3} using $\LS{D}{5/2}$-shelved detection,  4.3$\times$ lower than the best observed error using standard detection. For $\LS{F}{7/2}$-shelved detection with a 1~ms detection time, the detection error is reduced by another factor of $300\times$ ($12\times$) to \num{6(7)e-6} (\num{6.3(3)e-4}) on the APD (EMCCD).

\section{Conclusion}

In this work, we demonstrate that it is possible to combine the benefits of a long-lived, first-order magnetic-field insensitive hyperfine qubit with the high-fidelity detection typically observed in an optical qubit. By first shelving the population in one qubit state of $\Yb$ to a metastable level, we are able to use high-power, near-resonant Doppler cooling light to perform efficient state discrimination without suffering off-resonant leakage. To enable scaling to larger qubit registers, we also characterize the detection error after shelving to $\LS{D}{5/2}$ when using a spatially-resolving EMCCD. For both the APD and EMCCD detectors we compare the performance of software routines for processing photon detection data.  This involves either analyzing time-resolved information about the incoming photons collected on the APD, or exploiting spatial correlations between EMCCD pixels using a classifier routine. 

We also validate that the state detection error in our system can be further reduced by shelving to the long-lived $\LS{F}{7/2}$ manifold. New laser systems used in one of two schemes we outline could be used to overcome the prohibitively long optical pumping time to $\LS{F}{7/2}$, making this a viable avenue for ultra-high-fidelity detection.

Ultimately our results foreshadow the possibility of combining novel data-processing software routines with physics-based techniques in the future to further reduce measurement errors without requiring extensive hardware modifications. When combined with an efficient repump laser at 760~nm to reset the qubit state, we believe the electron-shelving based detection routine presented here will improve the practicality and scalability of current $\Yb$ quantum devices.

\section*{Acknowledgements}
This work was partially supported by the Intelligence Advanced Research Projects Activity Grant No. W911NF-16-1-0070, the US Army Research Office Grant No. W911NF-14-1-0682, the Australian Research Council Centre of Excellence for Engineered Quantum Systems Grant No. CE170100009, and a private grant from H.\&A. Harley.


%

\end{document}